# Integrated Reference Cavity for Dual-mode Optical Thermometry and Frequency Stabilization

QIANCHENG ZHAO,[1] MARK W. HARRINGTON,[1] ANDREI ISICHENKO,[1] RYAN O. BEHUNIN,[2,3] SCOTT B. PAPP,[4] PETER T. RAKICH,[5] CHAD W. HOYT,[6] CHAD FERTIG,[6] AND DANIEL J. BLUMENTHAL[1,*]

[1]*Department of Electrical and Computer Engineering, University of California, Santa Barbara, Santa Barbara, CA, 93106, USA*
[2]*Department of Applied Physics and Material Science, Northern Arizona University, Flagstaff, AZ, 86011, USA*
[3]*Center for Materials Interfaces in Research and Applications (¡MIRA!), Northern Arizona University, Flagstaff, AZ, 86011, USA*
[4]*Time and Frequency Division 688, National Institute of Standards and Technology, Boulder, CO, 80305, USA*
[5]*Department of Applied Physics, Yale University, New Haven, CT, 06511, USA*
[6]*Honeywell Aerospace, Plymouth, MN, 55441, USA*
*\*danb@ucsb.edu*

**Abstract:** Optical frequency stabilization is a critical component for precision scientific systems including quantum sensing, precision metrology, and atomic timekeeping. Ultra-high quality factor (Q) photonic integrated optical resonators are a prime candidate for reducing their size, weight and cost as well as moving these systems on chip. However, integrated resonators suffer from temperature-dependent resonance drift due to the large thermal response as well as sensitivity to external environmental perturbations. Suppression of the cavity resonance drift can be achieved using precision interrogation of the cavity temperature through the dual-mode optical thermometry. This approach enables measurement of the cavity temperature change by detecting the resonance difference shift between two polarization or optical frequency modes. Yet this approach has to date only been demonstrated in bulk-optic whispering gallery mode and fiber resonators. In this paper, we implement dual-mode optical thermometry using dual polarization modes in a silicon nitride waveguide resonator for the first time, to the best of our knowledge. The temperature responsivity and sensitivity of the dual-mode TE/TM resonance difference is 180.7±2.5 MHz/K and 82.56 μK, respectively, in a silicon nitride resonator with a $179.9 \times 10^6$ intrinsic TM mode Q factor and a $26.6 \times 10^6$ intrinsic TE mode Q factor. Frequency stabilization is demonstrated by locking a laser to the TM mode cavity resonance and applying the dual-mode resonance difference to a feedforward laser frequency drift correction circuit with a drift rate improvement to 0.31 kHz/s over the uncompensated 10.03 kHz/s drift rate. Allan deviation measurements with dual-mode feedforward-correction engaged shows that a fractional frequency instability of $9.6 \times 10^{-11}$ over 77 s can be achieved. These results show promises to realize dual-mode thermometry techniques in all-waveguide integrated resonators, opening the door to on-chip photonic integrated reference cavities and ultra-stable lasers.

## 1. Introduction

Optical reference cavities [1–7] play a critical role in laser linewidth narrowing and frequency stabilization for applications including optical communications [8–10], atomic clocks [11–14], spectroscopy [15,16], and quantum computation [17,18]. Frequency-stabilization techniques that incorporate Pound-Drever-Hall (PDH) locking [19] utilize high quality factor (Q) reference cavities with large optical mode volumes in vacuum housed single crystal silicon [3] or ultra-low expansion (ULE) glass cavities [20], to achieve better than $10^{-15}$ fractional frequency stability over a second, leading to sub-Hertz integral linewidths and ultra-stable operation. For low-cost and portable applications, it is desirable to miniaturize these cavities using photonic integrated waveguide-based designs. Optical microresonators are a good candidate due to their compact size and reduced mass, which can offer improved performance with respect to environmental disturbances. Bulk-optic resonators based on silica [4] and crystalline fluoride materials [1,2,21] are widely used due to their ultra-high Q factors and high finesse. Recently, bulk-optic vacuum cavities with lithographically defined silica mirrors have achieved an astounding finesse over 700,000 [22]. By leveraging recent breakthroughs in ultra-low-loss waveguide fabrication technique, all-waveguide silicon nitride ($Si_3N_4$) ring resonators have approached half-a-billion intrinsic Q factor [23], moving towards that of silica and crystalline cavities. These integrated all-waveguide resonators have the advantage of monolithic integration with other components to realize wafer-scale, CMOS compatible, planar lightwave circuits [24], opening the door to high-performance on-chip frequency stabilization.

A major challenge with waveguide-based integrated resonators is their increased sensitivity to thermal fluctuations compared to their vacuum spaced counterparts, due to the thermorefractive and thermal expansion effects in the cavity materials. Thermodynamic noise source can be mitigated by engineering the thermorefractive coefficients [25] or to suppress cavity thermal expansion using a low-expansion substrate [6]. Another approach is active control of the laser frequency using high sensitivity cavity temperature measurements through differential shifts in the cavity resonance. Dual-mode optical thermometry (DMOT) has been used to precisely probe the cavity temperature by detecting the difference in thermal responses between two polarization [26] or two frequency [27] modes. The measured intra-cavity temperature is used as an error signal for feedback control to stabilize the cavity temperature [28] or with feedforward control to regulate the laser frequency [29]. Advances in miniaturizing the DMOT stabilization technique include crystalline whispering gallery mode (WGM) resonators [30] and fiber resonators [31]. However, to the authors' knowledge, DMOT-based laser frequency stabilization has not yet been realized in a photonic integrated waveguide resonator.

We demonstrate, for the first time to our knowledge, an all-waveguide DMOT-based feedforward frequency drift stabilization scheme. The on-chip integrated optical reference cavity is made of ultra-low-loss $Si_3N_4$ waveguide that is designed to support high Q TE and TM modes, with intrinsic Q factors of $26.6 \times 10^6$ and $179.9 \times 10^6$, respectively. The cavity exhibits a resonance difference temperature responsivity of $180.7 \pm 2.5$ MHz/K which is over 3 times larger than a fiber resonator [31]. The temperature input sensitivity of the system is 82.56 μK that is 10 times smaller than a lithium niobate resonator [32]. The measured resonance difference between two polarizations is employed as an error signal to feed into a feedforward circuit. The feedforward circuit utilizes an acousto-optic modulator (AOM) to stabilize the laser whose frequency is PDH locked the TM resonance of the cavity. The drift-corrected laser shows 10 times smaller peak-to-peak frequency variations against external temperature perturbations. In the presence of external temperature ramping, the frequency drift rate is reduced to 0.31 kHz/s as compared to 10.32 kHz/s without stabilization, a 32 times reduction. The Allan deviation (ADEV) of the feedforward-corrected laser frequency reaches $9.6 \times 10^{-11}$ at 77 s.

The demonstrated DMOT-feedforward frequency drift correction scheme on an integrated $Si_3N_4$ resonator presents great potentials towards high-performance photonic integrated optical reference cavities and ultra-narrow linewidth lasers. An artistic illustration of an example photonic circuit that employs a DMOT reference cavity is shown in Fig. 1. An integrated laser is frequency locked to the DMOT reference cavity. The dual-mode resonance difference of the cavity is monitored by the electronic circuit. The circuit will correct the locked laser frequency via driving the cavity heater and/or tuning the optical frequency modulator, to yield a long-term drift-free narrow-linewidth laser frequency. Such a chip-scale solution will be useful to field deployed applications in precision metrology, quantum sensing and atomic clocks, and energy-efficient optical coherent communications.

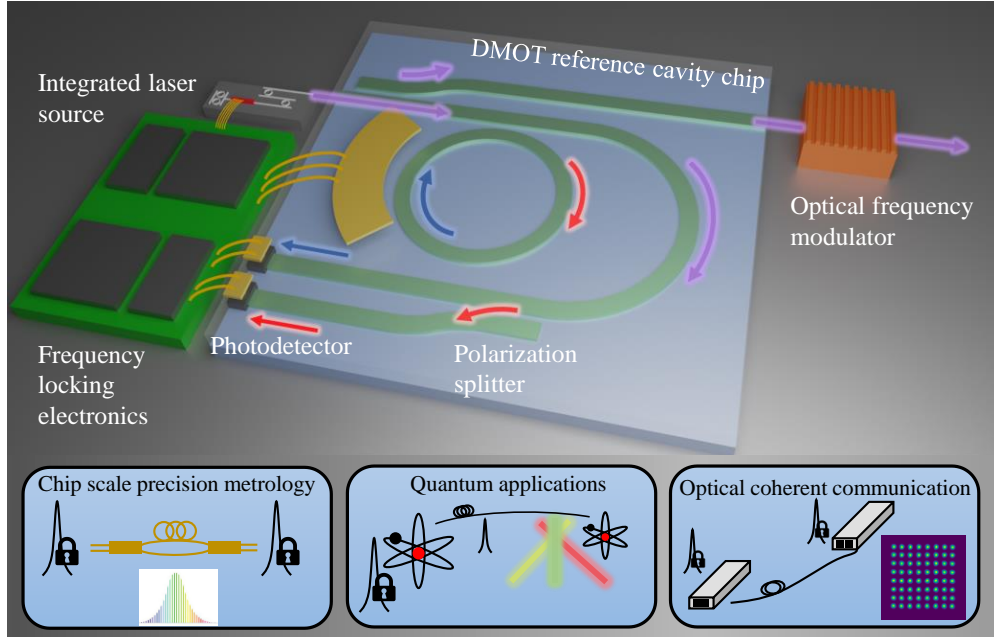

Fig. 1 Artistic illustration of a DMOT reference cavity with an integrated laser and frequency locking electronics. An integrated laser is frequency locked to the DMOT reference cavity. The dual-mode resonance difference of the cavity is monitored by the electronic circuit. The circuit will correct the locked laser frequency via driving the cavity heater and/or tuning the optical frequency modulator. The on-chip frequency stabilized hybrid integrated laser can be used in precision metrology, quantum and atomic applications, and energy-efficient optical coherent communications.

## 2. Results

The principle of DMOT in our multimode resonator is shown in Fig. 2(a). The waveguide has a cross-section of 80 nm × 6 μm that can support both TE and TM polarization modes. An all-pass ring resonator is designed with a radius of 8530.8 μm and a 3.5 μm ring-bus coupling gap. By manipulating the fiber polarization controller paddles, both TE and TM modes can be coupled into the waveguide. Due to the mode area and confinement difference, the TE and TM modes have unequal effective thermo-optic coefficients, leading to a temperature-dependent resonance difference which is a sensitive indicator of the local cavity temperature. Thus, the cavity temperature variation can be readout through the resonance difference change.

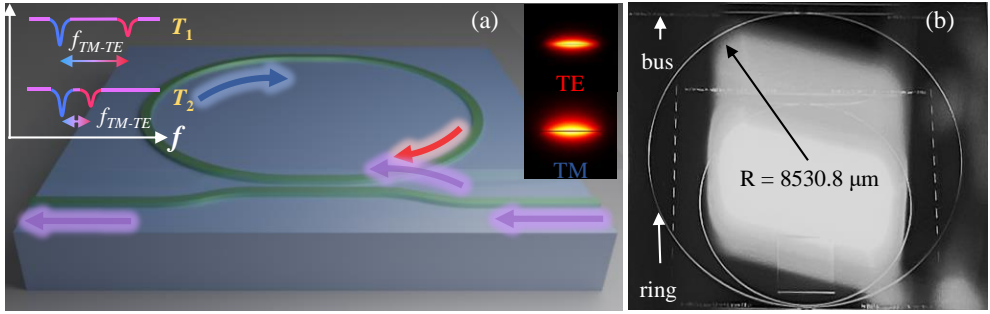

Fig. 2 (a) Artistic illustration of a dual-mode optical thermometry-based silicon nitride reference cavity. The red, blue, and purple colors represent the TE mode, TM mode, and mixer polarizations, respectively. The mode profiles of the two polarizations are plotted in the top-right corner. (b) A monochromatic contrast enhanced photo of the fabricated device. Some testing structures are also included in the same die including a smaller ring resonator.

## 2.1 Resonator characteristics

The device is fabricated using our ultra-low-loss $Si_3N_4$ waveguide fabrication technique [23], as shown in Fig. 2(b) (see Supplement 1 for details). The transmission spectrum of the fabricated resonator is shown in Fig. 3(c) illustrating the multimode nature of the resonator. The measured free spectral range (FSR) of the TM mode is 30.12 pm corresponding to 3.758 GHz at 1550 nm. The group index of the TM mode is $n_{g,TM} = 1.4883$. The FSR of the TE mode is 29.08 pm corresponding to 3.629 GHz, and its group index $n_{g,TE}$ is 1.5413 at 1550 nm. The loaded Q factors of the resonators are measured and verified using two different techniques: a radiofrequency calibrated Mach-Zehnder interferometer (MZI) method and a ring-down method (See Supplement 1 for more details). In the former approach, the full width at half maximum (FWHM) of the TM resonance (Fig. 3(a)) is $\Delta\nu_{TM} = 21.37$ MHz through Lorentzian curve fitting, giving a loaded Q factor $Q_{L,TM} = \nu_{res,TM} / \Delta\nu_{TM} = 9.05\times10^6$ at 1550 nm, where $\nu_{res,TM}$ is the TM mode resonant frequency. The FWHM of the TE mode (Fig. 3(b)) is $\Delta\nu_{TE} = 7.44$ MHz and its loaded Q factor $Q_{L,TE} = 25.99\times10^6$. The loaded Q factor of the TM mode is further confirmed through the ring-down method. The decay time ($\tau_{TM}$) is extracted to be 7.69 ns from the exponential curve fitting of the dissipated cavity power, and the loaded Q factor is measured to be $9.34\times10^6$, agreeing well with the results from the MZI method. The ring-down method was not applied to the TE mode because the TE modes have small extinction ratios that degrade the signal-to-noise (SNR) ratio when monitoring the dissipated cavity power. With the measured group index and resonance lineshape, the waveguide propagation loss of the TM and TE modes are calculated to be 0.15 dB/m and 1.02 dB/m, respectively. The TM mode has much smaller propagation loss than the TE mode, since the TM mode has a weaker mode confinement and thus a less scattering loss and waveguide material loss. The intrinsic Q factors of the TM and TE modes are $179.87\times10^6$ and $26.61\times10^6$, respectively. Due to the mismatches of mode areas and propagation losses of the TM and TE modes, the ring-bus coupling gap is carefully chosen so that both TM and TE resonances are supported. This non-optimum coupling condition leads to small extinction ratios for both resonances, since the TM resonance is over coupled and the TE resonance is under coupled. The characteristics of the TE and TM modes are summarized in Table 1.

Table 1. Summary of TE and TM resonance characteristics

| Mode | $Q_L$ (×10$^6$) | $Q_{int}$ (×10$^6$) | FWHM [MHz] | Extinction ratio | FSR [GHz] | $n_g$ | α [dB/m] |
|---|---|---|---|---|---|---|---|
| TM | 9.05 | 179.87 | 21.37 | 1.24 | 3.758 | 1.4883 | 0.15 |
| TE | 25.99 | 26.61 | 7.44 | 1.09 | 3.629 | 1.5413 | 1.02 |

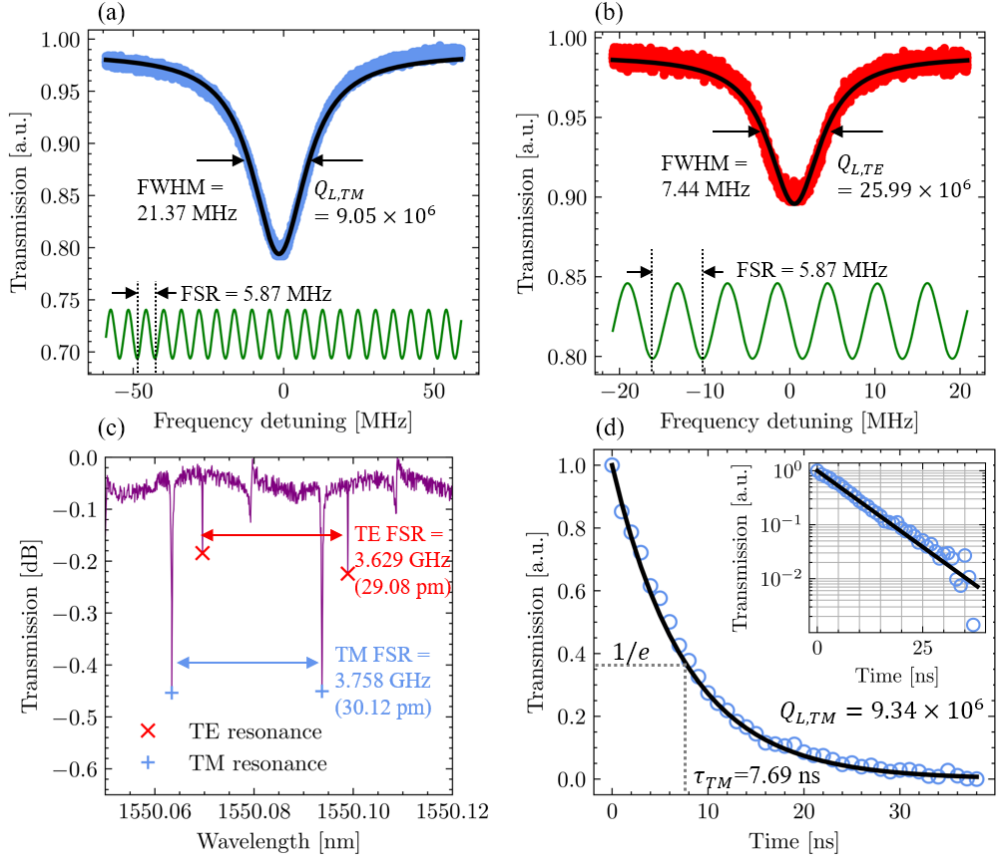

Fig. 3 Linewidth, FSR, ring-down measurements. (a) The transmission spectrum of the TM resonance at 1550 nm. The loaded Q factor is derived from its FWHM. An unbalanced MZI with an FSR of 5.871 ± 0.004 MHz (green curve) is used to calibrate the frequency detuning. (b) The transmission spectrum of the TE mode with its FWHM and the loaded Q factor denoted. (c) The FSR measurement of the TM and TE resonances. The resonances of the TM and TE modes are respectively denoted by the blue "+" and red "×" signs. The FSRs of the TM and TE modes are 3.758 GHz and 3.629 GHz, respectively. (d) Ring-down measurement of the TM resonance results in a cavity lifetime of 7.69 ns, corresponding to a $Q_L$ of 9.34 million. A first-order exponential decay function (black solid line) was fit to the normalized resonator transmission output (circle), and the time stamp corresponding to the 1/e point (where e $\approx$ 2.71828) is extracted as the TM mode ring-down time. The ring-down time, $\tau$, and the 1/e transmission are indicated by the gray dashed lines. The inset shows the transmission output (circle) in logarithmic scale, and the solid line shows the fitting of the data.

*2.2 Dual-mode resonance difference temperature responsivity*

The linearized temperature dependence of a resonant frequency, $f$, can be attributed to the thermal expansion and thermorefractive effects as described below for a dielectric waveguide resonator [33],

$$\frac{1}{f}\frac{df}{dT} + \frac{1}{n_g}\frac{dn_\text{eff}}{dT} + \frac{n_\text{eff}}{n_g}\alpha_E = 0 \qquad (1)$$

where $n_\text{eff}$ is the effective mode refractive index, $n_g$ is the group index, $\alpha_E$ is the linear thermal expansion coefficient, $dn_\text{eff}/dT$ is the effective thermo-optic coefficient of the waveguide mode. We assume that the two polarization modes have large spatial overlap, meaning $\alpha_E^\text{TE} \approx \alpha_E^\text{TM}$.

Here the superscripts indicate the TE and TM polarizations. Thus, the temperature dependence of the TM and TE resonance difference is

$$\frac{df_{\text{TM-TE}}}{dT} = -f\left[\left(\frac{1}{n_g^{\text{TM}}}\frac{dn_{\text{eff}}^{\text{TM}}}{dT} - \frac{1}{n_g^{\text{TE}}}\frac{dn_{\text{eff}}^{\text{TE}}}{dT}\right) + \left(\frac{n_{\text{eff}}^{\text{TM}}}{n_g^{\text{TM}}} - \frac{n_{\text{eff}}^{\text{TE}}}{n_g^{\text{TE}}}\right)\alpha_E\right] \quad (2)$$

Using the measured thermo-optic coefficients of $Si_3N_4$ and $SiO_2$ as $24.5\times10^{-6}$ /K [34] and $9.8\times10^{-6}$ /K [35] at 1550 nm, we simulated the $dn_{\text{eff}}^{\text{TM}}/dT$ to be $9.861\times10^{-6}$ /K and $dn_{\text{eff}}^{\text{TE}}/dT$ to be $11.204\times10^{-6}$ /K, respectively. Thus, the corresponding TM/TE resonance difference temperature responsivity, denoted by $df_{\text{TM-TE}}/dT$, is 152.7 MHz/K. Since the optical mode is weakly guided and a large portion of the optical power propagates in the $SiO_2$ cladding, it is reasonable to use the linear thermal expansion coefficient of amorphous $SiO_2$, which is $0.56\times10^{-6}$ /K [36], as an approximation for the true linear thermal expansion coefficient of the composite waveguide. Thus, we can estimate the absolute resonance temperature responsivity. For example, the TM mode resonance temperature responsivity, $df_{\text{TM}}/dT$, is calculated to be 1387.1 MHz/K based on Eq. (1). The ratio of the TM resonance temperature responsivity to the dual-mode resonance difference temperature responsivity is 9.08 with the chosen parameters.

The temperature responsivity of the dual-mode resonance difference of the fabricated resonator is characterized by measuring its resonance differences of the two polarization modes from the cavity transmission spectra at different temperatures, as shown in Fig. 4(a). Each spectrum is normalized and offset at its measurement temperature for clarity. As the spectra are aligned to the TM resonances, one can find that the TE resonances move linearly with respect to the temperature change, as indicated by the black dashed line. The frequency differences between these two orthogonal modes are plotted as a function of the temperature in Fig. 4(b), and the slope gives $df_{\text{TM-TE}}/dT =187.56$ MHz/K. The measured responsivity matches reasonably well with the theoretical value. The mismatch comes from the difference of the material properties and waveguide geometry between the theoretical analysis and the real devices.

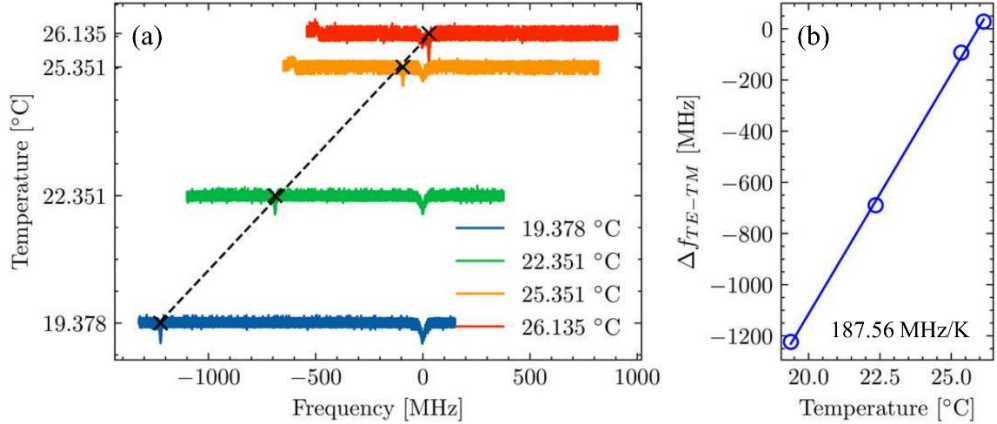

Fig. 4 (a) transmission spectra at various temperatures. The spectra are individually horizontally translated to align TM resonances at zero detuning frequency, and are vertically offset at their corresponding temperatures for clarity. The black "x" denotes the TE resonance position. The TE/TM resonance difference as a function of temperature is plotted in (b). The slope of the linear fitted line corresponds to the measured temperature responsivity

### 2.3 Double-sideband Pound-Drever-Hall frequency locking system

With the obtained dual-mode resonance difference temperature responsivity, the cavity temperature change can be inferred by probing the dual-mode resonance difference. To read out the dual-mode resonance difference, a DSB PDH system is built as shown in Fig. 5(a). A semiconductor diode laser (L) to be stabilized is frequency locked to one of the TM mode cavity resonances forming the first PDH loop (blue path). A voltage-controlled oscillator (DSB VCO) that drives the electro-optic modulator (EOM) is tuned and frequency locked to the TE/TM frequency difference such that the TE sideband created by the EOM aligns with TM carrier frequency, forming the second PDH loop (red path). When both PDH loops are enabled, the laser frequency is locked a TM resonance of the cavity and the DSB VCO frequency is locked to the dual-mode resonance difference. To illustrate the frequency locking positions, the TM and TE mode transmission spectra and their error signals are depicted in Fig. 5(b), and their measurement positions are denoted in Fig. 5(a) as (1), (2), (3), and (4). The purple line in Fig. 5(b) marks the locked laser frequency.

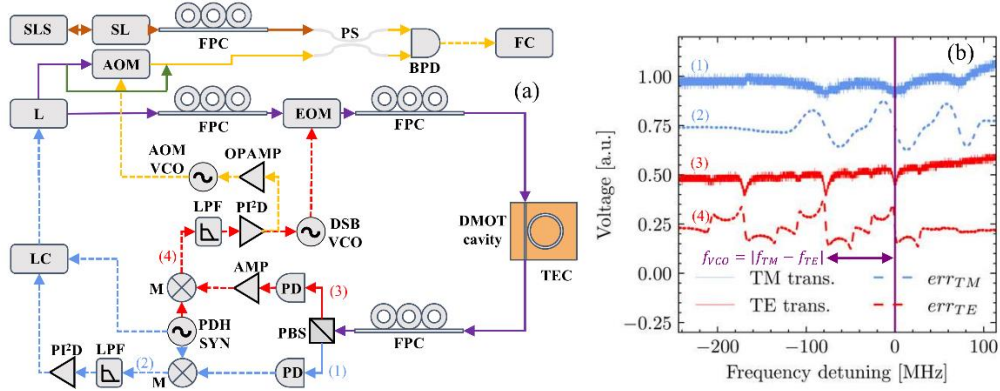

Fig. 5 (a) Dual-mode optical thermometry feedforward stabilization system diagram. The solid lines represent optical paths, and the dashed lines represent electrical paths. The blue color is related with the optical TM mode, and the red color is related with the optical TE mode. The purple path includes both TE and TM modes. The yellow path is used for feedforward stabilization. The green solid line is the optical path that bypasses the AOM. The AOM is bypassed in the temperature responsivity experiment but is engaged in the frequency stabilization experiment. **L**: laser; **LC**: laser controller; **PI²D**: proportional-integral-differential controller; **LPF**: low-pass filter; **M**: mixer; **PDH SYN**: RF synthesizer for PDH locking; **PD**: photodetector; **PBS**: polarization beam splitter; **FPC**: fiber polarization controller; **VCO**: voltage controlled oscillator; **TEC**: thermoelectric cooler; **AMP**: RF amplifier; **OPAMP**: operational amplifier; **FC**: frequency counter; **EOM**: electro-optic modulator; **AOM**: acousto-optic modulator; **SLS**: Stable Laser System reference cavity; **PS**: 50:50 power splitter; **BPD**: balanced photodetector. (b) TM (blue) and TE (red) transmission spectra (solid lines) and their PDH error signals (dashed lines). The probing locations of the measured signals are labeled as (1), (2), (3) and (4) in (a). The purple solid line shows the locked laser frequency position.

To measure the frequency drift of the DSB-PDH-locked laser, a fraction of the laser power is photomixed with another stabilized laser (SL) that refers to a ultra-stable Fabry-Perot cavity made of ULE glass (HV-6020-4, Stable Laser Systems, Boulder, CO). The SLS cavity has a linewidth of 1.55 kHz and < 5 kHz/day frequency drift at 1550 nm which is negligible in our experiments. The frequency drift of the DSB-PDH-locked laser is corrected through a feedforward configuration in which the laser passes through an AOM, and its output frequency is shifted by the AOM RF frequency. Highlighted as the yellow path in Fig. 5(a), we split the error signal that drives the DSB VCO into an operational amplifier (OPAMP) with adjustable gain, and use the output voltage to control the AOM VCO which drives the AOM. By setting a proper feedforward gain, the change of the AOM RF frequency is equal and in opposite sign to the change of the laser frequency. Thus, the AOM output frequency is compensated and corrected.

*2.4 Relation between dual-mode resonance difference and laser frequency change*

The DSB PDH frequency locking system allows us to monitor the dual-mode resonance difference as well as the absolute laser frequency change. In a DMOT feedforward frequency stabilization system, the dual-mode resonance difference is used to interrogate the cavity temperature variation which causes the absolute laser frequency change. To use the dual-mode resonance difference as an error signal to correct the laser frequency drift through an AOM, a linear correlation between the dual-mode resonance difference and the laser frequency is necessary for the feedforward configuration.

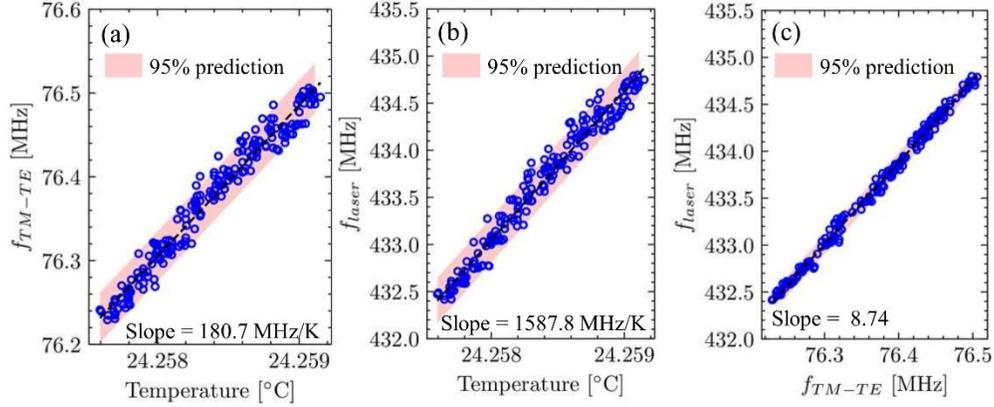

Fig. 6 (a) Dual-mode resonance differences vs. the TEC temperature. (b) Laser frequency vs. the TEC temperature. (c) The laser frequency vs. dual-mode resonance difference. The slopes of the linear-fitted black dashed lines in (a) and (b) are the temperature responsivities. The 95% predication interval zone is highlighted in pink.

To investigate the relation between the dual-mode resonance difference and the laser frequency, we change the cavity temperature slowly through a thermoelectric cooler (TEC) to make sure the TEC and the cavity temperature reaches thermal equilibrium. The dual-mode resonance difference is readout by a frequency counter and plotted as a function of the TEC temperature in Fig. 6(a). A linear curve fitting reveals its temperature dependence is 180.7±2.5 MHz/K, consistent with the transmission spectra measurement. The standard deviation of the $f_{TM\text{-}TE}$ is 14.9 kHz, corresponding to a measured temperature input sensitivity of 82.56 μK. The laser frequency is also plotted with respect to the TEC temperature shown in Fig. 6(b). Its temperature responsivity is linearly fitted to be 1587.8±19.9 MHz/K, agreeing reasonably well with our theoretical predictions from Eq. (1) within 13% mismatch. Combining the data from Fig. 6(a) and Fig. 6(b), the laser frequency is compared to the dual-mode resonance difference in Fig. 6(c), where a linear relation can be observed and the slope is derived to be 8.74±0.05, matching well with simulations within 4% difference. It should be noted that the feedforward path in Fig. 5(a) was not enabled in this experiment, because we need the temperature-driven laser frequency variation to extract its thermal responsivity. The laser light bypasses the AOM as indicated by the green path in Fig. 5(a).

*2.5 Laser frequency drift compensation*

To demonstrate the frequency stabilization against environmental temperature disturbance, we modulated the TEC temperature using a square wave with an amplitude of 2 mK as an external perturbation. The red dashed lines in Fig. 7 are the temperature setpoint curves. Without feedforward stabilization (Fig. 7(a)), $\Delta f_{\text{laser}}$ follows $\Delta f_{\text{TM-TE}}$ with a square wave-like profile. The peak-to-peak variation of $\Delta f_{\text{laser}}$ is 6.96 MHz in a duration of 480 s. When feedforward stabilization is engaged (Fig. 7(b)), the peak-to-peak variation of $\Delta f_{\text{laser}}$ is suppressed 10 times smaller to be 0.64 MHz in the same time duration. The overshoot in the frequency curves is caused by the temperature overshoot which is determined by the TEC temperature PID controller parameters.

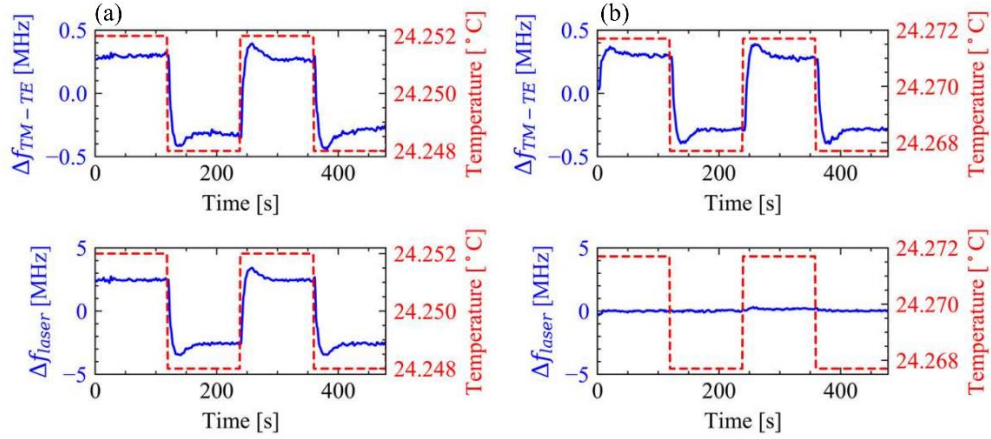

Fig. 7 Laser frequency variation against temperature perturbation. (a) Without feedforward stabilization, the dual-mode resonance difference change, $\Delta f_{TM\text{-}TE}$ (upper), and the laser frequency change, $\Delta f_{laser}$ (lower), are plotted along with the external temperature perturbations (red). (b) With feedforward stabilization engaged, the change of the laser frequency is greatly suppressed.

The slow drift compensation capability is further demonstrated by comparing the time series data of the laser frequency before and after the feedforward loop is engaged. To exaggerate the contrast, we added environmental perturbation by programming the TEC temperature setpoint linearly. The $\Delta f_{laser}$ starts initially with a frequency drift rate of -10.03 kHz/s in the first 240 s. For comparison, the scaled $\Delta f_{TM\text{-}TE}$ (green trace in Fig. 8(a)) is overlaid upon $\Delta f_{laser}$ with a scaling factor of 8.74. Before feedforward stabilization is engaged, both $\Delta f_{laser}$ and $\Delta f_{TM\text{-}TE}$ drift over time proportionally. Once the feedforward correction is applied, the laser frequency drift rate is reduced to -0.31 kHz/s as indicated by the red dashed line in Fig. 8(a), while the $\Delta f_{TM\text{-}TE}$ keeps increasing over time. A zoomed-in trace of the stabilized laser frequency revealing the details is shown in the inset. The Allan deviations (ADEV) of the laser frequencies with and without feedforward stabilization are compared in Fig. 8(b). Despite the external temperature perturbations, the feedforward-corrected laser frequency can still reach an ADEV of $9.6\times10^{-11}$ at 77 s, whereas the ADEV of the unstabilized laser frequency is as high as $3\times10^{-9}$ at the same integration time.

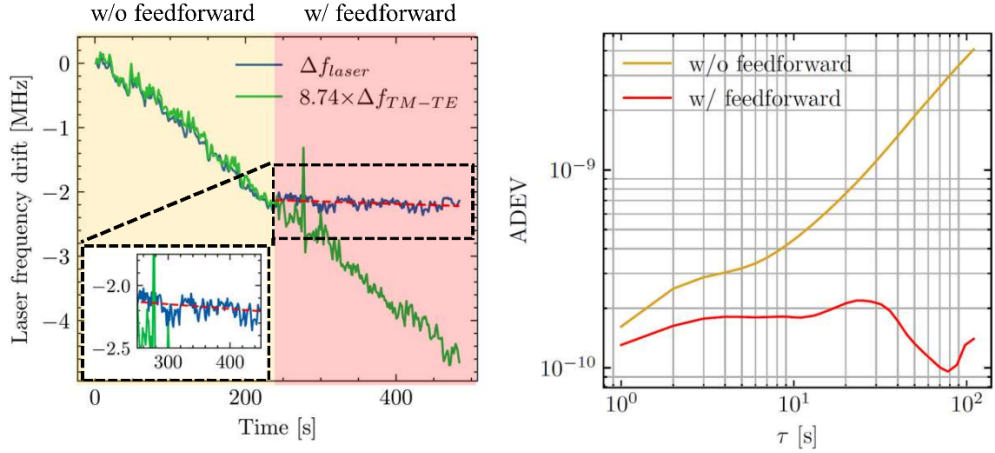

Fig. 8 (a) Measurement of the laser frequency drift (blue) and resonance difference drift (green, scaled) with and without feedforward stabilization when the chip temperature is ramped. The feedforward stabilization is engaged at 240 s. The dashed red line represents the linear curve fitting of the feedforward corrected frequency. The zoomed-in plot of the stabilized laser frequency is shown in the inset. (b) Comparison of the ADEVs of the laser fractional frequencies with and without feedforward stabilization when the chip temperature is ramped.

## 3. Discussion

We have demonstrated an all-waveguide $Si_3N_4$ reference cavity based on dual-mode optical thermometry (DMOT) feedforward stabilization in an ambient environment. The reference cavity has ultra-low propagation losses for the TE and TM modes. The resonator utilizes the differential thermal responses from the TE and TM modes while suppressing the common noises to probe the cavity temperature. We calculate that there is a 63.16% mode overlap that contributes to common mode cavity noise suppression. We measure the temperature responsivity to be 180.7±2.5 MHz/K, more responsive than that of a fiber resonator [31]. Its temperature input sensitivity is 82.56 μK, nearly 10 times smaller than a lithium niobate resonator [32], enabling a precision on-chip sensing of cavity temperature change. The linear correlation between the dual-mode resonance difference and laser frequency enables laser stabilization using a feedforward control circuit. We compare the laser frequency stability in the presence of external temperature perturbations with and without feedforward stabilization where the laser frequency drift is suppressed over 30 times with stabilization applied over without stabilization.

Table 2 Comparison of the DMOT device and system performance

| Form factor | Material | $Q_{L,TM}$ (x$10^6$) | $Q_{L,TE}$ (x $10^6$) | Responsivity (d$f_{TM-TE}$/d$T$) [MHz/K] | d$f_{laser}$/d$f_{TM-TE}$ | Temperature sensitivity | Environment | Ref. |
|---|---|---|---|---|---|---|---|---|
| Integrated | $Si_3N_4$ | 9.05 | 25.99 | 180.7 | 8.74 | 82.56 μK | Ambient | This work |
| WGM | $MgF_2$ | 2100 | 2100 | 78.71 | 21.66 | 8.53 μK | Vacuum | [30] |
| Fiber | silica | 170 | 170 | 48.3 | 33.3 | 70 nK | Ambient | [31] |
| WGM | $MgF_2$ | 2000 | 1000 | 89.8 | 19.3 | 480 nk | Ambient | [37] |
| WGM | $LiNbO_3$ | 0.2 | 0.29 | 834 | | 0.8 mK | Ambient | [32] |

| | | | | | | | |
|---|---|---|---|---|---|---|---|
| WGM | MgF$_2$ | 390 | 390 | 170 | 17.2 | 480 nK | Vacuum | [38] |

We summarize the performance of our DMOT stabilization system with representative published works (Table 2). Our resonator has a higher d$f_{TM-TE}$/d$T$ compared to MgF$_2$ and silica resonators, indicating more responsivity to cavity temperature change. Compared to WGM resonators and fiber resonators, the all-waveguide integrated resonator can be monolithically integrated on chip without fiber or prism couplings to cavities, thus reducing the mechanical and environmental coupling noises.

This work focuses on demonstrating dual-mode frequency stabilization in an all-waveguide device, and as such is not yet intended to perform as well as table-top bulk-optic and fiber systems. However, we see a path forward for improvements including improving temperature sensitivity and suppressing technical noises such as vibrations and intensity noise. The sensitivity of the dual-mode temperature sensor is ultimately set by the minimum detectable separation between the dual resonances which is determined by the cavity linewidths of the two resonances. To further improve the temperature sensitivity, resonators with higher loaded Q factors for both polarizations will be needed. The temperature sensitivity can be further improved by embedding this dual-mode sensor into an integrated stimulated Brillouin scattering (SBS) laser [7] where fiber versions have demonstrated < 100 nK temperature resolution [29]. With recent progress in ultra-high-Q microresonators (> 422 Million Q [23]), there is a path forward to integrate an SBS laser and dual-mode reference cavity into a single chip.

**Funding.**

Defense Advanced Research Projects Agency (FA9453-19-C-0030) and Advanced Research Projects Agency-Energy (DE-AR0001042). The views and conclusions contained in this document are those of the authors and should not be interpreted as representing official policies of DARPA, ARPA-E or the U.S. Government or any agency thereof.

**Acknowledgments.**

The authors would like to thank Naijun Jin from Yale University for constructive discussions on dual-mode optical thermometry system modeling, and Akshar Jain from University of California, Santa Barbara, for providing test equipment. A portion of the work was performed in the UCSB Nanofabrication Facility, an open access laboratory. Andrei Isichenko acknowledges the support from the National Defense Science and Engineering Graduate (NDSEG) Fellowship Program.

**Disclosures.**

The authors declare no conflicts of interest.

**Data availability.**

The data that support the plots within this paper and other findings of this study are available from the corresponding author on reasonable requests.

**Supplemental document.**
See Supplement 1 for supporting content.

# Integrated Reference Cavity for Dual-mode Optical Thermometry and Frequency Stabilization: supplemental document

This document provides supplementary information to "Integrated Reference Cavity for Dual-mode Optical Thermometry and Frequency Stabilization". In section 1, we present the device fabrication procedure. In section 2, we show the loaded Q factor measurement method using a Mach-Zehnder interferometer and a tunable laser. In section 3, we explain the steps to measure the loaded Q factor through a ring-down method.

**1. Device fabrication methods.**

An 80 nm-thick stoichiometric $Si_3N_4$ film was deposited using low-pressure chemical vapor deposition (LPCVD) method on a 100 mm-diameter silicon wafer with a pre-grown 15 μm-thick thermal oxide. The wafer was spun coated with standard deep ultraviolet (DUV) anti-reflective and photoresist layers and then patterned using an ASML PAS 550/300 DUV stepper. The photoresist layer is developed by the MIF300 developer and the anti-reflective layer is etched by $O_2$ plasma in a PlasmaThem etching tool. The waveguide patterns were transferred to the $Si_3N_4$ layer through dry etching in a Panasonic E640 inductively coupled plasma etcher using a $CHF_3/CH_4/O_2$ recipe. The residual photoresist patterns were stripped by $O_2$ ashing in a Panasonic E626I inductively coupled plasma tool, followed by soaking in hot N-methyl-2pyrrolidone solution at 80°C with ultrasonic bath and rinsing in isopropanol. The organic impurities were further removed by dipping the wafer in a freshly prepared piranha solution heated at 110°C. An additional plasma clean using a Gasonics Aura 2000-LL Downstream asher tool helped remove the leftover organic residuals and moisture. A layer of 15 nm $Si_3N_4$ was deposited on the etched wafer using a LPCVD tool. Then the wafer went through wet oxidation for 30 minutes at 1100°C, which could convert the 15 nm extra $Si_3N_4$ layer into silicon dioxide, bringing the $Si_3N_4$ layer thickness back to its original design. A 6 μm tetraethyl orthosilicate (TEOS)-precursor $SiO_2$ was deposited as the waveguide upper cladding using a plasma-enhanced chemical vapor deposition (PECVD) tool. The deposited wafer was annealed at 1050°C for 7 hours and 1150°C for 2 hours. The fabricated wafer then underwent dicing into several die for experimental characterization.

**2. Q-factor measurement using an RF calibrated MZI.**

A tunable laser (Velocity TLB-6730) is driven by an external function generator (< 100 Hz) for frequency detuning. A fraction of the frequency detuned laser power is passed through an unbalanced MZI that has a calibrated free spectral range of 5.871 ± 0.004 MHz. The transmitted power from the MZI is monitored on a synchronized oscilloscope. The rest of the laser power is injected into the resonator, and the transmitted power is overlaid on the oscilloscope as well. By simultaneously scanning the laser frequency through both the MZI and the device-under-test, the MZI fringe patterns provide a radiofrequency calibrated frequency reference for accurate evaluation of resonance linewidth. The FWHM of the resonator is estimated by fitting the resonator transmission spectrum to a Lorentzian curve.

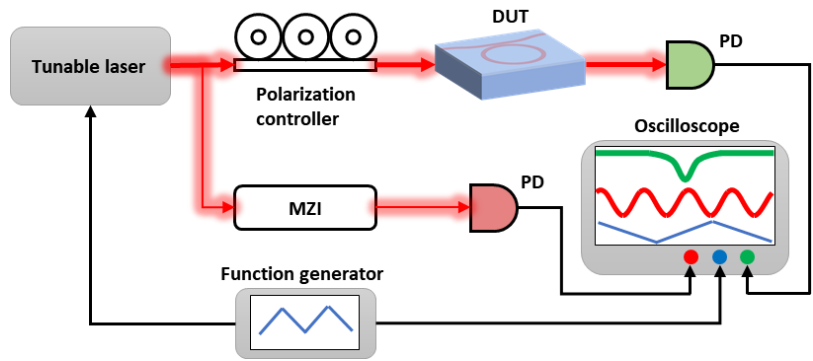

Fig. S 1 A schematic is shown for the RF calibrated MZI setup. The function generator outputs a sawtooth waveform to sweep the tunable laser frequency. 10% of the laser power goes to an unbalanced MZI, while 90% of the power is injected into the cavity. The transmission spectrum of the MZI provides a reference spectrum to measure the frequency detuning because its fringe patterns have been pre-calibrated to be 5.871 MHz.

### 3. Q-factor measurement using ring-down.

A tunable laser (Velocity TLB-6730) is frequency swept around a resonance by applying a triangular voltage input to the laser piezoelectric transducer. The laser intensity is controlled by an intensity modulator (Avanex SD20) which is driven by a 50% duty cycle, 10 kHz square wave RF signal with switching time < 10 ns. The square wave is synchronized such that when the laser frequency is at cavity resonant frequency, the input optical power shuts down, and the optical power inside the cavity dissipates with an exponential decay. The decay time ($\tau$) was measured by monitoring the optical power from the cavity transmission port on an oscilloscope. The loaded Q-factor can be calculated using the equation $Q_L = \omega_{res} \tau$, where $\omega_{res}$ is the resonant angular frequency.